\documentclass[12pt]{iopart}

\usepackage{gensymb} 
\usepackage{graphicx}
\usepackage{xcolor}
\begin{document}

\title[FERMILAB-PUB-20-011-DI-LDRD-TD (accepted copy)]{First demonstration of a cryocooler conduction cooled superconducting radiofrequency cavity operating at practical cw accelerating gradients}

\author{R.C. Dhuley, S. Posen, M.I. Geelhoed, O. Prokofiev, and J.C.T. Thangaraj}

\address{Fermi National Accelerator Laboratory, Batavia, Illinois 60510, USA}
\ead{rdhuley@fnal.gov}
\vspace{10pt}

\begin{abstract}
We demonstrate practical accelerating gradients on a superconducting radiofrequency (SRF) accelerator cavity with cryocooler conduction cooling, a cooling technique that does not involve the complexities of the conventional liquid helium bath. A design is first presented that enables conduction cooling an elliptical-cell SRF cavity. Implementing this design, a single cell 650 MHz Nb$_3$Sn cavity coupled using high purity aluminum thermal links to a 4 K pulse tube cryocooler generated accelerating gradients up to 6.6 MV/m at 100\% duty cycle. The experiments were carried out with the cavity-cryocooler assembly in a simple vacuum vessel, completely free of circulating liquid cryogens. We anticipate that this cryocooling technique will make the SRF technology accessible to interested accelerator researchers who lack access to full-stack helium cryogenic systems. Furthermore, the technique can lead to SRF based compact sources of high average power electron beams for environmental protection and industrial applications. A concept of such an SRF compact accelerator is presented.
\end{abstract}

%
%
%
%
%

\section{Introduction}

Electron irradiation is a proven technique for environmental protection applications such as the treatment of industrial/municipal wastewater, flue gases, sewage sludge, etc. and has been demonstrated on several pilot scale projects~\cite{Chmielewski2011}. For electron irradiation to be competitive on the large scale with existing treatment methods, electron beam (e-beam) sources capable of providing beam energy of \mbox{1$-$10 MeV}, very high average beam power, and high wall-plug efficiency ($>$50\%) are needed~\cite{DOE2015}. The sources must also be robust, reliable, and have turn-key operation to be viable in the harsh environment expected around these applications~\cite{DOE2015}. Compact sources with smaller footprints and lower infrastructure cost are also preferred. 

E-beam sources using superconducting radiofrequency (SRF) cavities as the beam accelerator can meet several of the above requirements. A meter-long or even a shorter structure of standard niobium cavities~\cite{Grassellino2017} or of low-dissipation Nb$_3$Sn cavities~\cite{Posen2015}, both of which easily generate accelerating gradients \mbox{$>$10 MV/m}, can be an electron source with the desired beam energy. The low surface resistance of SRF cavities reduces their surface losses and provides high efficiency transfer of the input RF power to the beam, which can help to achieve the wall-plug efficiency target. The low surface resistance also facilitates constructing cavities with a larger aperture and allows RF operation with 100\% duty cycle (continuous wave or cw mode), both of which are favorable for generating and efficiently transporting beams of very high average power. SRF cavities, however, need operation at cryogenic temperatures and are conventionally cooled by immersion in baths of liquid helium held near \mbox{2$-$4.5 K}. The cryogenic infrastructure~\cite{White2014} needed for compressing, liquefying, distributing, recovering, and storing helium as well as expert cryogenic operators~\cite{NagimovIPAC} needed for oversight run counter to the robustness, high reliability, compactness, and turn-key operation desired in industrial settings. 

An approach to simplify the helium cryogenic infrastructure and reduce its footprint is to integrate a closed-cycle \mbox{4 K} cryocooler into an SRF cryomodule and recondense in-situ the boil-off helium gas produced by the cavity dynamic heat dissipation. Although this compact and operationally simpler cooling scheme, as implemented at the JAERI FEL~\cite{KikuzavaSRF}, was shown to work reliably over year-long periods, it still relies on a liquid helium bath, leading to some undesirable requirements: (1) a separate helium cryosystem/liquid inventory for initially filling the cryomodule, (2) rigorous pressure vessel and relief design of the cryomodule as it contains a bath of liquid helium, and (3) large helium gas compressors and a storage system to recover the helium during warm-up.

Conduction cooling an SRF cavity by directly connecting to a closed-cycle cryocooler with a thermally conductive link will eliminate the need for the conventional helium bath. This elimination leads to dramatic simplification of the accelerator: (1) a liquid helium inventory, a helium recovery/storage system, and a helium pressure vessel and relief design is no longer needed, (2) the cryogenics becomes very reliable (commercial \mbox{4 K} cryocoolers have mean time between maintenance of \mbox{$>$20000 hrs} (\mbox{2.3 years})~\cite{Cryomech}), safe (no liquid helium safety and oxygen deficiency hazards), and simple to operate (cryocoolers turn on/off with push of a button), and (3) significantly reduced footprint as well as added option of portability because all of the cryogenics is integrated into the cryomodule. Following its conceptualization~\cite{Kephart2015} in 2015, conduction cooling of SRF cavities has been studied albeit only by means of computer simulations. Previous work is limited to understanding its feasibility based on multiphysics (electromagnetic and thermal) simulations~\cite{Holzbauer2014,Kostin2018} and a design of an e-beam accelerator using a conduction cooled SRF cavity~\cite{Ciovati2018}. A program to demonstrate practical accelerating gradients on conduction cooled SRF cavities began at Fermilab in 2016. In this letter we present experimental results from this program, demonstrating cw accelerating gradients up to \mbox{6.6 MV/m} on a single cell SRF cavity. 

\section{Development of conduction cooling for elliptical SRF cavities}

\begin{figure}
\includegraphics[scale=0.45]{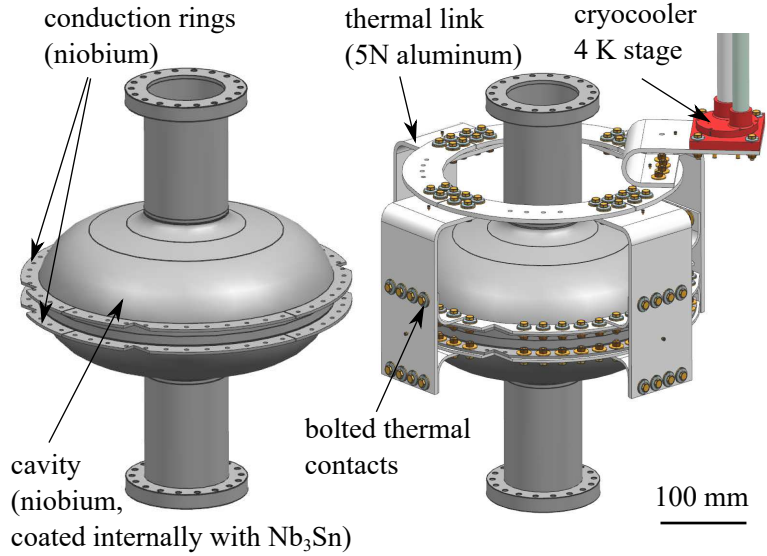}
\caption{\label{fig:cavity} Design of a single cell 650 MHz niobium cavity for conduction-cooling: niobium rings electron-beam welded to the cavity near its equator (left) and a 5N aluminum thermal link~\cite{Dhuley2019IEEE} connecting the cavity to the \mbox{4 K} stage of a cryocooler (right). The bolted thermal contacts were prepared as described in our prior report~\cite{Dhuley2018Cryo}. The cavity was coated internally with a $\sim$2 $\mu$m thick Nb$_3$Sn layer before cryogenic testing.}
\end{figure}

The elliptical single-cell niobium cavity used for the this work has the following parameters: resonance frequency \mbox{650 MHz}, accelerating length, $L_{acc}$ = \mbox{0.23 m}, shape factor, $G$ = \mbox{265 $\Omega$}, and normalized shunt impedance, $r/Q$ = \mbox{152 $\Omega$}. For conduction cooling, niobium rings (SRF grade, RRR$>$300) were e-beam welded to the two elliptical half-cells as illustrated in Figure~\ref{fig:cavity}. The cavity surface was prepared by removing \mbox{120 $\mu$m} via electropolishing (EP), 3 hour \mbox{800 $\degree$C} vacuum furnace treatment, \mbox{20 $\mu$m} light EP, and high pressure rinse with water. After initial performance evaluation, the cavity inner surface was coated with a $\sim$2 $\mu$m thick layer of Nb$_3$Sn, grown via vapor diffusion~\cite{Posen2017}, to enable low dissipation operation~\cite{Posen2015} near \mbox{4.5 K}. The cavity was then cooled in \mbox{4.4 K} liquid helium in the Fermilab Vertical Test Stand (VTS) to obtain a baseline of quality factor, $Q_0$ $vs.$ cw accelerating gradient, $E_{acc}$. The cavity was then warmed, removed from the VTS, and prepared for conduction cooling without disturbing the inner vacuum.

A thermal conduction link~\cite{Dhuley2019IEEE} of 5N aluminum (purity $>$99.999\%) was machined out of stock plates, cleaned to remove surface oxide, and bolted to the cavity niobium rings. The bolting procedure~\cite{Dhuley2018Cryo} involves interposing a \mbox{4 mil} thick foil of indium between the niobium and aluminum plates and pressing the contact with \mbox{2 kN} force applied by a silicon bronze screw, a brass nut, and stainless steel Belleville disc springs. The other termination of the thermal link was bolted to the \mbox{4 K} stage of a pulse tube cryocooler. The cavity-thermal link assembly was then installed on a test setup~\cite{Dhuley2019CEC} (conduction-cooled test setup or CCTS) recently developed at Fermilab. This setup is comprised of a vacuum vessel, a magnetic shield (an enclosure with \mbox{$\sim$10 mG} background), a thermal radiation shield, and a Cryomech PT420 two-stage pulse tube cryocooler (rated to provide cooling of \mbox{2 W} @ \mbox{4.2 K} with \mbox{55 W} @ \mbox{45 K}). A new RF power source was also developed that can supply \mbox{10 W} @ \mbox{650 MHz} of cw power to the cavity, measuring the forward, reflected and transmitted powers, and locking the source frequency to the instantaneous resonance frequency of the cavity. For recording temperature of the cavity-cryocooler assembly, the cavity carried four cryogenic thermometers affixed to the niobium rings and the cryocooler carried one cryogenic thermometer on its \mbox{4 K} stage. The cavity temperature referred to in this letter is the average of the four cavity thermometer readings.

\section{Cavity performance}

\begin{figure}[!t]
\includegraphics[scale=0.75]{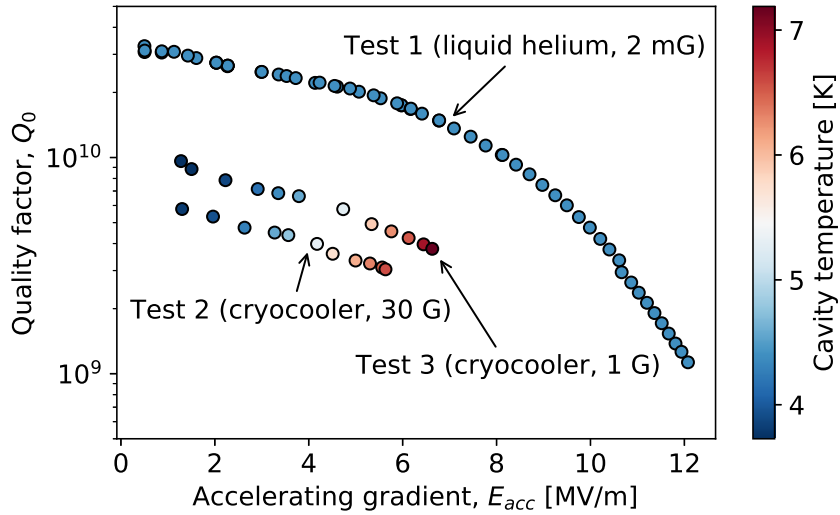}
\caption{\label{fig:Q0Eacc} Cavity quality factor, $Q_0$ $vs.$ accelerating gradient, $E_{acc}$ measured on a single cell, Nb$_3$Sn coated, \mbox{650 MHz} niobium SRF cavity. The data uncertainty is $<$10\%. Test 1 used a helium bath cooled cavity at \mbox{4.4 K}, in Fermilab Vertical Test Stand with background magnetic field of \mbox{2 mG}. In Tests 2 and 3, the cavity was conduction-cooled with a \mbox{2 W} @ \mbox{4.2 K} pulse tube cryocooler. The improvement in Test 3 resulted from the reduction of magnetic field around the cavity when magnetic disc springs (residual \mbox{$\sim$30 G}) were replaced with cleaner disc springs (residual \mbox{$<$1 G}).}
\end{figure}

Three RF tests were performed including one with liquid helium (baseline) and two with cryocooler conduction cooling. Figure~\ref{fig:Q0Eacc} shows the cavity quality factor, $Q_0$ $vs.$ cw accelerating gradient, $E_{acc}$ (both accurate to within 10\%), determined using standard cavity measurement procedure~\cite{Melnychuk2014}. Test 1 was carried out in the Fermilab VTS with liquid helium and witnessed carefully controlled conditions $viz.$ a background magnetic field of \mbox{$\sim$2 mG} and slow/uniform cooldown with rate of \mbox{0.1 K/min} through the Nb$_3$Sn superconducting transition temperature~\cite{Peiniger1988} of \mbox{18 K}. Both these factors reduce the residual surface resistance of Nb$_3$Sn, which enhances the $Q_0$ of the cavity~\cite{Posen2017}. During the RF measurements, the cooling power of the helium bath was regulated using a vapor pumping system so that the cavity remained isothermal at \mbox{$\sim$4.4 K} over the range of $E_{acc}$. Test 1 recorded $Q_0$ of 3x10$^{10}$ at $E_{acc}$ of \mbox{1 MV/m} and $Q_0$ of 4x10$^9$ at $E_{acc}$ of \mbox{10 MV/m}. The highest gradient of \mbox{$\sim$12 MV/m} recorded in Test 1 was limited by RF power. 

For Test 2, the cavity was cooled conductively using the cryocooler to below \mbox{4 K}, with a slow cooldown rate of \mbox{0.03 K/min} through the Nb$_3$Sn transition temperature. Although the magnetic shield of the CCTS provided a background of \mbox{$\sim$10 mG}, we later found that some stainless-steel disc springs on the thermal link had residual field as high as 30 G. The slow cooldown in such high magnetic field is expected to trap the flux in the Nb$_3$Sn layer, causing the cavity $Q_0$ to degrade significantly. A $Q_0$ of 6x10$^9$ at $E_{acc}$ of \mbox{1 MV/m} was recorded in Test 2, which is five times smaller than in Test 1. Limited by the power output of the RF source, the cavity produced maximum $E_{acc}$ of \mbox{$\sim$5.5 MV/m} during Test 2. For Test 3, magnetically cleaner disc springs with residual of \mbox{$<$1 G} were installed on the thermal link. The cavity showed noticeable improvement: $Q_0$ of 10$^{10}$ was measured at $E_{acc}$ of \mbox{1 MV/m} and the cavity sustained maximum $E_{acc}$ of \mbox{$\sim$6.6 MV/m}, limited again by input RF power. 

We note in Figure~\ref{fig:Q0Eacc}, a distinction between the $Q_0$ $vs.$ $E_{acc}$ data measured with liquid helium and cryocooler conduction cooling. As previously mentioned, the helium bath temperature control system in the VTS (Test 1) held the cavity isothermal over the range of $E_{acc}$, yielding a $Q_0$ $vs.$ $E_{acc}$ curve at the near-constant temperature of \mbox{$\sim$4.4 K}. In the CCTS, however, there was no temperature regulation system on the cryocooler. So as heat dissipation in the cavity increased with the increase in $E_{acc}$, the steady state temperature of the cavity increased as well. The color gradient in the data for Test 2 and Test 3 reflects this effect. Thus, unlike Test 1, the $Q_0$ $vs.$ $E_{acc}$ data from Test 2 and Test 3 do not correspond to a fixed cavity temperature but rather have the cavity temperature vary from \mbox{$\sim$4 K} to \mbox{$\sim$7 K} depending on the $E_{acc}$.  

\begin{figure}
\includegraphics[scale=0.6]{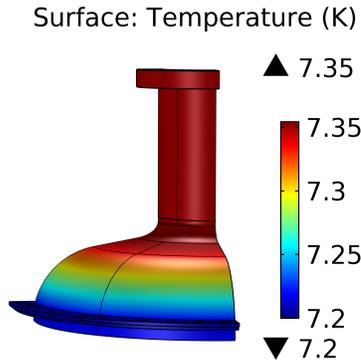}
\caption{\label{fig:CavTemp} Simulated cavity surface temperature during operation at \mbox{6.6 MV/m} $E_{acc}$. The conduction ring inner faces are held at the measured temperature of 7.2 K and 4 W of RF heat is applied on the inner surface of the cavity. The temperature difference between the ring and the cavity iris is \mbox{$\sim$0.15 K}.}
\end{figure}

To understand the effectiveness of conductive heat extraction from locally near the equator, we plot in Figure~\ref{fig:CavTemp} the simulated surface temperature of the cavity during operation at \mbox{6.6 MV/m} $E_{acc}$. The simulation uses the following boundary conditions: (a) temperature of \mbox{7.2 K} at the inner faces of the conduction rings and (b) \mbox{4 W} of RF heat applied on the cavity inner surface. Both of these were measured/determined from the experiments. Due to its small thickness and strong adhesion, the Nb$_3$Sn layer is assumed to be in local thermal equilibrium with the niobium substrate. The temperature dependent thermal conductivity of niobium is estimated using the procedure given in~\cite{Bonin1996SUST}, using RRR=300 for the niobium. The temperature difference between the ring and the cavity iris is \mbox{$\sim$0.15 K}, indicating good thermal uniformity over the cavity surface. The overall temperature remains much below the superconducting transition temperature of Nb$_3$Sn. The cooling technique can therefore be considered effective, especially because it is demonstrated on a \mbox{650 MHz} cavity, which is large in size compared to other common elliptical SRF cavities (1.3 GHz, 3.9 GHz). 

\begin{figure}[!t]
\includegraphics[scale=0.75]{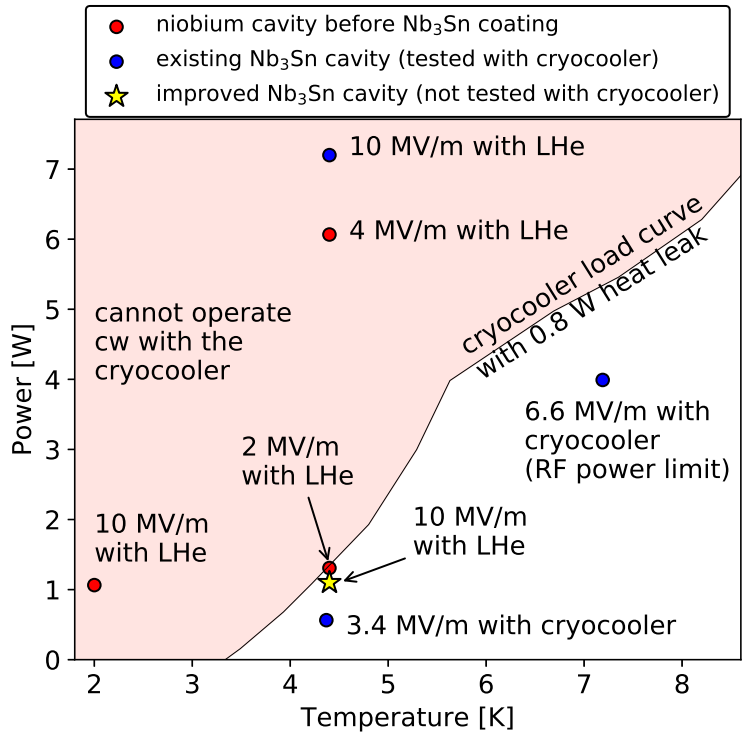}
\caption{\label{fig:PowerTemp} Dissipated power $vs.$ temperature for the \mbox{650 MHz} Nb$_3$Sn single cell cavity under study compared with the PT420 cryocooler load curve. The load curve accounts for the \mbox{0.8 W} heat leak prevailing during measurements. Observed notable cw gradients with conduction cooling (with their current limits) marked on the graph are: \mbox{3.4 MV/m} at \mbox{$\sim$4.4 K} (cooling capacity) and \mbox{6.6 MV/m} at \mbox{$\sim$7.2 K} (RF power). A $Q_0$ of $\sim$3x10$^{10}$ at \mbox{4.4 K} is needed to achieve \mbox{10 MV/m} with conduction cooling. An improved Nb$_3$Sn coating has already attained this $Q_0$ in a recent test~\cite{PosenSRF2019} in the Fermilab VTS. For comparison, we show representative gradients measured in the VTS on the cavity before Nb$_3$Sn coating.}
\end{figure}

Figure~\ref{fig:PowerTemp} presents a summary of the present findings in terms of the cavity temperature, dissipated power, and the corresponding $E_{acc}$. The plot is divided into two regions by the cryocooler load curve, accounted for the \mbox{0.8 W} heat leak prevailing during the measurements. cw operation is not possible with the cryocooler in the shaded region because here the dissipated power exceeds the cryocooler capacity at a given temperature. For instance, the operation at \mbox{10 MV/m} cw at \mbox{4.4 K} with \mbox{$\sim$7.2 W} of dissipation lies in this region. The unshaded region allows cw operation with the cryocooler. At \mbox{4.4 K}, conduction cooling produced a modest $E_{acc}$ of \mbox{$\sim$3.4 MV/m}, limited by the cryocooling capacity at this temperature as well as due to the degraded $Q_0$ from flux trapping. However, with the increase in the cryocooling capacity with temperature, the cavity at \mbox{$\sim$7.2 K} generated an $E_{acc}$ of \mbox{$\sim$6.6 MV/m}. This suggests that the attainable $E_{acc}$ is not limited by the cryocooler cooling capacity at \mbox{$\sim$4.4 K} and significantly larger $E_{acc}$ can be generated by letting the system operate warmer than $\sim$4.4 K. 

Figure~\ref{fig:PowerTemp} also highlights that reaching practical cw gradients with a niobium cavity may not be feasible with the cryocooler. We show representative gradients obtained in the VTS on the cavity before coating with Nb$_3$Sn. $E_{acc}$ only upto 2 MV/m lay within the range of the cryocooler capacity at \mbox{4.4 K}. Achieving higher gradients either needed more cooling capacity (\mbox{$\sim$6 W} at \mbox{4.4 K} to reach \mbox{4 MV/m}) or higher $Q_0$ operation at colder temperature (\mbox{2 K} to produce \mbox{10 MV/m} cw), both of which are out of the cryocooler cooling range.

The $E_{acc}$ of \mbox{$\sim$6.6 MV/m} over $L_{acc}$ = \mbox{0.23 m} equals a calculated energy gain of \mbox{$\sim$1.5 MeV}. This clearly makes our existing configuration of one-cell cavity with one-cryocooler practicable for treatment of industrial flue gas~\cite{Ciovati2018}. The attainable $E_{acc}$ with one cryocooler can be pushed up by improving the $Q_0$ of our cavity. The ongoing efforts for Nb$_3$Sn coating optimization have already produced a $Q_0$ of $\sim$3x10$^{10}$ at \mbox{10 MV/m} cw on a similar \mbox{650 MHz} single-cell cavity~\cite{PosenSRF2019} in the Fermilab VTS. The corresponding dissipation of \mbox{$\sim$1.1 W} at \mbox{4.4 K} is now in the regime of cryocooler conduction cooling as marked in Figure~\ref{fig:PowerTemp}. Replicating this performance with conduction cooling requires improvements to the magnetic hygiene of our CCTS. These improvements are underway including complete replacement of stainless-steel disc springs with those made of non-magnetic beryllium-copper.

\section{Outlook and future work}

\begin{figure}[!b]
\includegraphics[scale=0.6]{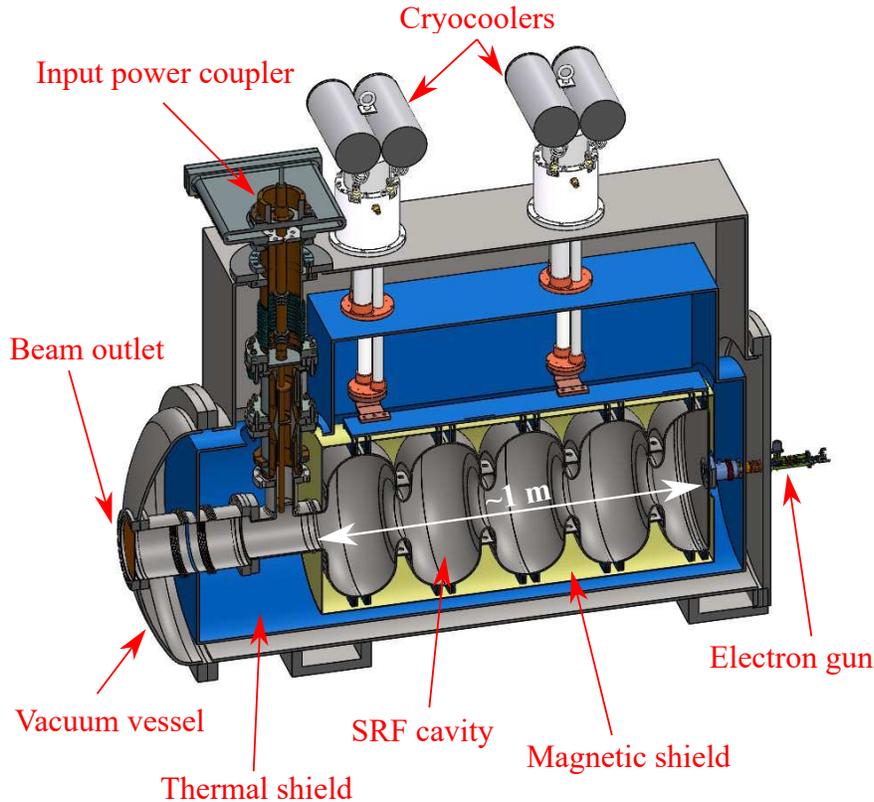}
\caption{\label{fig:Accelerator} CAD rendering of a compact, meter-long e-beam SRF accelerator. The accelerator has a 4.5-cell 650 MHz cavity conductively coupled to the 4 K stage of four 2 W cryocoolers. Intermediate cooling for the thermal radiation shield is provided by the 50 K stage of the cryocoolers. The accelerator also has a low heat leak input RF power coupler~\cite{Kazakov2019} and a cavity-integreated electron gun~\cite{Mohsen2019}.}
\end{figure}

Large accelerators (example, LCLS-II~\cite{Ravindranath2017}) using hundreds of cavities require kilowatt-level refrigeration at liquid helium temperatures. We emphasize that cryocoolers are not economical for such large-scale cooling demand due to their lower efficiency: a large helium cryoplant requires \mbox{$\sim$0.4--0.8 kW(electrical)/W(cooling)} while a cryocooler requires \mbox{$>$10 kW(electrical)/W(cooling)}. However, it can be an enabler for a new class of compact, small-scale SRF accelerators, a concept of which is illustrated in Figure~\ref{fig:Accelerator}. Here we envision a 10 MeV e-beam source comprising of a meter-long 5-cell \mbox{650 MHz} SRF cavity generating \mbox{10 MV/m} cw~\footnote{The Nb$_3$Sn SRF program at Fermilab has produced a 9-cell 1.3 GHz Nb$_3$Sn cavity that generated \mbox{$>$10 MV/m} cw in the Fermilab VTS~\cite{PosenSRF2019}. This is a critical step towards producing multi-cell SRF cavities needed for industrial SRF accelerators described in this letter.}. With \mbox{$\sim$6--7 W} of dissipation at \mbox{4.4 K}, the cavity can be conduction-cooled using four 2 W cryocoolers. The design for components suitable for such a machine $viz.$ a low heat leak input RF power coupler~\cite{Kazakov2019} and an cavity-integrated electron source~\cite{Mohsen2019} is currently underway.   

We introduced a new method to cool an SRF cavity to cryogenic temperatures by conductively coupling to a closed-cycle 4 K cryocooler. The method when adopted in an SRF accelerator will eliminate the conventional liquid helium bath and offer robustness, reliability, and turn-key cryogenic operation, potentially making the SRF accelerator attractive for industrial settings. A \mbox{650 MHz} Nb$_3$Sn single-cell cavity generated cw gradient of \mbox{$\sim$6.6 MV/m} (calculated electron energy gain of \mbox{1.5 MeV}) with $Q_0$ of 4x10$^9$ when cooled using a \mbox{2 W} @ \mbox{4.2 K} pulse tube cryocooler. Continued work targets to improve $Q_0$ to push the $E_{acc}$ to \mbox{10 MV/m}, develop conduction-cooling for multi-cell SRF cavities, study potential cavity microphonics resulting from cryocooler vibration, and develop a compact SRF accelerator as a source for \mbox{1$-$10 MeV} energy, high average power e-beams for industrial and environmental applications. 

In parallel to our efforts, it is encouraging to witness the growing interest in conduction-cooling of SRF cavities as demonstrated on a smaller 1.5 GHz cavity by the Jefferson Lab SRF group~\cite{Ciovati2020}. 

\section{Acknowledgement}

This manuscript has been authored by Fermi Research Alliance, LLC under Contract No. DE-AC02-07CH11359 with the U.S. Department of Energy, Office of Science, Office of High Energy Physics. The work was supported by Fermilab Laboratory Directed Research and Development (LDRD). We sincerely thank Dr. Robert D. Kephart, scientist emeritus at Fermilab, for his foundational work on this concept. We also thank the staff of Illinois Accelerator Research Center, Fermilab Accelerator Division Mechanical Support, and Fermilab SRF group for several technical contributions to this work. The Nb$_3$Sn coating infrastructure was supported by Fermilab LDRD and S. Posen DOE Early Career Award. We thank Dr. Roman Kostin and Yubin Zhao of Euclid Techlabs, LLC for their assistance with the cavity simulations and the design of the RF system used in the present work.

\section{References}

\bibliography{CCSRF}

\end{document}